\documentclass[10pt,conference]{IEEEtran}
\usepackage{epsfig,setspace,amsmath,epsf,amssymb,bm,theorem,cite,graphicx, epstopdf,float,color,mathtools,authblk,physics,algorithm,algorithmic}
\usepackage{optidef}
\usepackage{booktabs}
\usepackage{subfigure}
\usepackage{bbm}
\usepackage{tikz}
\usetikzlibrary{positioning,arrows.meta,calc,shapes.geometric}

\definecolor{axisblue}{RGB}{100,180,230}
\tikzset{
  aoii line/.style = {dashed, thick, axisblue},
  drop/.style      = {thick, black},
  dot/.style       = {circle, fill=black, inner sep=0pt, minimum size=6pt},
  reddia/.style    = {diamond, fill=red!70!black, inner sep=0pt,
                      minimum size=7pt},}

\IEEEoverridecommandlockouts
\allowdisplaybreaks

\begin{document}


\title{Minimizing Age of Incorrect Information in Pull-Based Remote Estimation of a Markov Source\thanks{I. Cosandal and S. Ulukus are with the University of Maryland, College Park, MD, USA. N. Akar is with Bilkent University, Ankara, T\"{u}rkiye. Corresponding author: S. Ulukus (email: ulukus@umd.edu).

This work is done when N.~Akar was on sabbatical leave as a visiting professor at the University of Maryland, MD, USA.

The material in this paper was presented in part at the IEEE ICC 2025.}}

\author{Ismail Cosandal \qquad Sennur Ulukus \qquad Nail Akar}

\maketitle
\begin{abstract}
    In this paper, we consider a discrete-time pull-based remote estimation system involving an information source and a remote monitor where the information source is a finite-state irreducible discrete-time Markov chain (DTMC) process. In this system, the remote monitor sends a pull request towards the source which responds by sampling its process so as to transmit the sampled values towards the monitor through a delay channel, encapsulated in information packets. We consider i) zero delay (ZD) and ii) unit delay (UD) channels, where the packet transmission times are zero and one slot, respectively. For remote estimation, we use the maximum a posteriori (MAP) estimator, and for quantifying its performance, we use the mean age of incorrect information (MAoII) which is a recently proposed freshness and mismatch metric that penalizes an incorrect estimation along with its duration. Our goal is to obtain pull policies so as to minimize MAoII under a long-run sampling rate constraint. For the ZD model, we propose a semi-Markov decision process (SMDP) formulation of the problem, which allows us to find the optimum observation-dependent pull policy. On the other hand, the UD channel prevents the monitor from possessing perfect information on the actual AoII process at any given time, and the SMDP formulation of the ZD channel cannot readily be used for the UD model. For the system with the UD channel, we first provide a numerical algorithm to calculate the MAoII and the average sampling rate for a given observation-dependent pull policy. For the optimization of the underlying remote estimation system, we use the joint distribution of the age and source processes to be obtained from the history of all observations, termed as {\em belief}. Subsequently, we propose a belief-dependent policy obtained via the belief-MDP formulation and a constrained deep reinforcement learning (CDRL) algorithm.
\end{abstract}

\section{Introduction}
Age of information (AoI) metric has recently been proposed to capture information freshness in remote estimation problems \cite{Yates__HowOftenShouldone}. The AoI metric quantifies information freshness from the perspective of a monitor that keeps track of how long ago the latest received information packet in the system had been generated. However, it is argued in \cite{maatouk2020} that AoI may fall short of capturing freshness in certain estimation problems since it does not consider the dynamics of the sampled process. For instance, even though the latest received packet may have been generated a long time ago, it is possible that the source may not have changed since then, and therefore, the packet can still be fresh. Similarly, a recently received packet may contain stale information if the source has already changed its state after the packet was generated. Stemming from these drawbacks of AoI, \cite{maatouk2020} proposes an alternative freshness metric, namely age of incorrect information (AoII), or in short age, that penalizes the mismatch between the source and its estimation over time, and regardless of when it is sampled, it defines the estimation as \emph{fresh} if it is the same as the source. Since AoII takes into account the dynamics of the source process, it is considered to be a semantic metric \cite{lu2022semantics, maatouk2022age} and is suitable for goal-oriented communications. Another interesting feature of AoII in contrast to AoI is that the monitor is not required to get a new sample to bring the age down to zero since the mismatch condition between the source and the monitor may as well be brought to an end with a transition of the source to the estimated value at the monitor.

\begin{figure}[t]
    \centering
    \begin{tikzpicture}[scale=0.28]
    \draw[rounded corners,thick,darkgray] (-1,-1) rectangle (7,4) {};
    \filldraw (3,2) node[anchor=center] {\small{monitor process}};
    \filldraw (3,0.75) node[anchor=center] {\small{$\hat X_t$}};
    \draw[rounded corners,thick,darkgray] (21.5,-1) rectangle (29,4) {};
    \filldraw (25.25,2) node[anchor=center] {\small{source process}};
    \filldraw (25.25,0.75) node[anchor=center] {\small{${X}_t$}};
    \filldraw (10.5,-0.5) node[anchor=north] {\small{$o_t$}};
    \draw[ultra thick] (7,2) -- (9,2);
    \draw[ultra thick] (9,2) -- (11,4);
    \draw[->,red,ultra thick] (11,4) -- (13,2);
    \node at (11,2) (nodeA) {};
    \node at (16,7) (nodeB) {};
    \draw[ultra thick,->] (13,2) -- (21.5,2);
    \filldraw (13.5,4.5) node[anchor=center] {\small{$a_t=0,1$}};
    \node at (8,-1) (nodeA) {};
    \node at (12,3) (nodeB) {};
    \draw[ultra thick,<-] (7,-0.45) -- (12.25,-0.45);
    \draw[rounded corners,thick,darkgray] (12.25,-1.95) rectangle (16.25,.95) {};
    \filldraw (14.25,-0.45) node[anchor=center] {\small{channel}};
   \draw[ultra thick,<-] (16.25,-0.45) -- (21.5,-0.45);
    \end{tikzpicture}
	\caption{Pull-based remote estimation system involving a source and a remote monitor.}
    \label{fig:sys}
\end{figure}
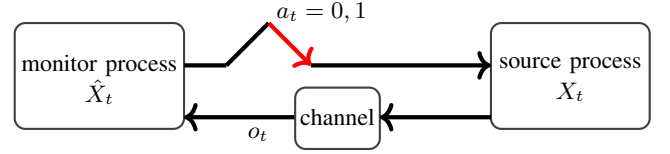

In this paper, we consider the pull-based remote estimation system in Fig.~\ref{fig:sys} involving an information source which is a finite-state irreducible discrete-time Markov chain (DTMC) process, denoted by $X_t, t \geq 0$, and a remote monitor which maintains an estimate $\hat{X}_t, t \geq 0$, for the original process $X_t$, from the updates received by the monitor from the source. At time $t$, the remote monitor either sends a pull request, i.e., $a_t=1$, to the source, which responds by sampling its process and transmitting the sample value towards the monitor. Alternatively, the monitor does not initiate a pull request, i.e., $a_t=0$. We consider two different channels, namely i) zero delay (ZD), ii) unit delay (UD) channels, where the packet transmission times are zero and one slot, respectively. In Fig.~\ref{fig:sys}, $o_t$ represents the observation, i.e., $o_t=X_t$ and $o_t=X_{t-1}$, at time $t$, if a packet is received for the ZD and UD channels, respectively. 
The case of channel errors and more general packet transmission times are left for future research. We propose to use the maximum a posteriori (MAP) estimator for both channels in addition to existing martingale estimators \cite{akar_ulukus_tcom24} that have been used in the majority of existing remote estimation works. For quantifying the performance of the estimator, we use the mean age of incorrect information (MAoII), which is the time average of the corresponding AoII process. Our goal is to obtain pull policies, i.e., when to pull, so as to minimize MAoII under a long-run sampling rate (i.e., pull-rate) constraint for both the ZD and UD channels.

For the ZD channel, we propose an average-cost constrained semi-Markov decision process (SMDP) formulation of the problem when the considered pull policy is a deterministic observation-dependent policy, i.e., the time to wait until the next pull instant is $g(j)$ slots if a pull request is made at time $t$ with corresponding observation $o_t=j$, i.e., $X_t =j$. SMDP is a generalization of MDP (see \cite{book_howard}) in which the time between two decision instants is generally distributed rather than a single slot, for all states and actions, and has been studied extensively in the literature \cite{book_ross, book_tijms, book_ibe}. We first solve the unconstrained SMDP problem with a proper selection of the cost function using policy iteration \cite{book_white,book_ibe}, and then use the Lagrangian method adapted from \cite{makowski1986} to solve the constrained SMDP problem so that the MAoII is minimized while the long-run pull-rate is kept below a desired value. Thanks to the resetting property of the AoII process upon packet transmissions, it is possible to cast the optimization problem for the ZD channel as an SMDP for which well-established solution methods exist in the literature.

On the other hand, the UD channel impedes the monitor from obtaining perfect information on the AoII process and the state of the original process, and hence the SMDP formulation cannot be used. For the UD channel, we first provide a numerical algorithm to exactly calculate the MAoII and the average sampling rate for an observation-dependent pull policy. This algorithm allows us to obtain the performance of the system for several benchmark policies, including the observation-dependent policy obtained from the ZD channel, which, however, is a sub-optimum policy for UD. Since we know neither the original process nor the age for the UD system, we resort to the use of {\em belief}, which stands for the a posteriori distribution of the age and source processes to be obtained from the history of all past actions and observations using Bayes' rule. We then formulate the problem as a belief-MDP (see \cite{smallwood1973optimal},\cite{pineau2003point}) which is a reformulation of a partially observable MDP (POMDP) into a fully observable MDP by ensuring that decisions are made based on this belief. However, the continuous nature of the belief renders it infeasible for the use of standard dynamic programming techniques. Therefore, we propose a belief-dependent policy obtained via the belief-MDP formulation using a constrained deep reinforcement learning (CDRL) algorithm. To solve the corresponding CDRL, we utilize the \emph{state-augmented} approach proposed in \cite{naderializadeh2022state} that guarantees the constraints are satisfied almost surely, and the resulting policy is near-optimal. We note that a similar approach has been taken in \cite{zakeri2024semantic}, where an optimal policy for an error-free channel is obtained through the MDP formulation, whereas DRL is used for the erasure channel counterpart. Another similar approach is taken in \cite{de2025minimizing} for a push-based energy harvesting problem.  

Our contributions are summarized as follows.
\begin{itemize}
\item Most of the existing work in remote estimation uses martingale estimators due to their simplicity and amenability to analysis. However, martingale estimators are inferior to MAP estimators for pull-based remote estimation. In this paper, we incorporate MAP estimation in our optimization framework.
    \item For the remote estimation system with the zero-delay channel, we develop optimum observation-dependent policies for minimizing MAoII under sampling rate constraints, for a general finite-state DTMC source, using the SMDP framework. To the best of our knowledge, the optimum solution is proposed for the first time in this general setting.
    \item  Using the theory of absorbing Markov chains (AMC), we develop a numerical algorithm to calculate the MAoII and the sampling rate for the remote estimation system with the unit delay channel model for a given observation-dependent pull policy. 
    \item For the optimization of the unit delay channel model, we propose a belief-dependent policy obtained via the belief-MDP formulation and CDRL algorithm. Since the optimum solution for this problem is not available, we demonstrate the effectiveness of the proposed approach by comparisons with various benchmark policies.
\end{itemize}

The paper is organized as follows. In Section~\ref{sec:related}, we present the related work. In Section~\ref{sec:prelim}, a brief introduction to semi-Markov decision processes and absorbing Markov chains is given, which is needed to follow the paper. In Section~\ref{sec:system}, we detail the system model. Sections~\ref{sec:zerodelay} and \ref{sec:unitdelay} are dedicated to the analysis and proposed solutions for the zero-delay and unit-delay channel models, respectively. Numerical results are presented in Section~\ref{sec:Numerical} to validate the effectiveness of the proposed approach. Finally, we conclude in Section~\ref{sec:conclusions}.
\section{Related Work}
\label{sec:related}

AoII is mainly investigated for symmetric or binary Markov chains in earlier studies, for which a single-threshold policy has been proposed to minimize the average AoII \cite{maatouk2020, chen2021minimizing, kriouile2022pull, zakeri2024semantic}. Additionally, in these works, the latest received information is used as the estimate, termed as the \emph{martingale estimator} in \cite{akar_ulukus_tcom24}, since it would be optimum only if the source were a martingale (see \cite{ross1996stochastic}). On the other hand, in our previous works \cite{cosandal2024modelingC, cosandal2024aoiiC, cosandal_etal_TRIT24}, we have shown for general asymmetric Markov chain source processes that a single-threshold policy would not be guaranteed to be optimum. More specifically, in \cite{cosandal2024aoiiC, cosandal_etal_TRIT24}, it was shown that the optimum transmission policy should take into account all the estimation, source, and age values. Similarly, it was shown in \cite{cosandal2024modelingC} that the monitor can reduce the average AoII when the pull request rates are allowed to be dependent on the latest received information. However, this particular work considers a preemption mechanism at the transmitter which aborts the transmission of the current information packet if the source process changes before the packet is received. Similarly, the pull-based AoII minimization problem in \cite{kriouile2022pull} considers an immediate transmission that allows the monitor to keep track of AoII. On the other hand, for the UD channel model of the current paper, the decision maker, i.e., the monitor, does not possess perfect information either on the source process or the AoII process.

AoII minimization problem in the presence of multiple sources and a centralized scheduler is investigated for pull-based settings in \cite{kriouile2022pull, raj2025design, ornee2025remote, cosandal2025sensor}. The works \cite{kriouile2022pull, raj2025design, ornee2025remote} study independent source processes, whereas \cite{cosandal2025sensor} considers a scenario involving multiple correlated sub-processes which evolve according to a single super-process, but each sensor only observes a single sub-process. Furthermore, ALOHA-like protocols are studied to minimize AoII in \cite{munari2024monitoring, chiariotti2025peak}.

POMDP is an extension of MDP \cite{Krishnamurthy_2016} where the action-taker cannot directly observe the state of the problem, but it can estimate the likelihood of the state from partial observations, i.e., belief, and the policy is defined as a function of the belief \cite{smallwood1973optimal, spaan2012partially}. In this formulation, the underlying process governing state transitions is a Markov process, and state transition probabilities and the cost function depend only on the state and the action. The study \cite{araya2010pomdp} further extends POMDP to address problems of belief-dependent cost, namely $\rho$POMDP, and demonstrates that if the cost function is convex, existing POMDP solutions can be adapted for $\rho$POMDP as well. In addition, the study \cite{spaan2015decision} considers a scenario in which an agent aims to maintain an accurate belief alongside its main goal, and it defines this type of problem as \emph{POMDP with informational rewards} (POMDP-IR). In this work, the authors propose expanding the action space with the estimation of the agent, and the cost function is modified to penalize/reward the incorrect/correct information. Additionally, in \cite{satsangi2018exploiting}, it is shown that these two formulations can be converted to each other.

The state space of the belief-MDP, which is termed as \emph{belief space}, contains uncountably many elements, and therefore it is not possible to obtain the value function for each belief state by using dynamic programming from the belief-MDP representation. One method to cope with the continuous belief space is the discretization of the belief space \cite{cregg2024reinforcement}, which, however suffers from low accuracy and high complexity for high and low discretization step sizes, respectively \cite{hauskrecht2003linear}. In some cases \cite{liu2023optimizing, liu2023optimizing2}, only a finite number of beliefs can be visited; hence, it is possible to convert the belief-MDP to an MDP with a finite state space, which does not apply to the problem studied in the current paper. Although an uncountable number of belief values can be visited in the current paper's setting, the number of possible next beliefs from an individual belief in a finite horizon is countable thanks to the finite-dimensional observation and action spaces. DRL has recently gained popularity for solving MDPs using the exploration-exploitation trade-off \cite{arulkumaran2017deep}. Indeed, the value function for a POMDP can be approximated from the belief using DRL \cite{egorov2015deep}, or from the observation \cite{jaakkola1994reinforcement}, and the action that minimizes the value function is applied as a sub-optimal policy.

POMDP formulation has been used to minimize AoI in several works \cite{shao2021partially, tahir2024collaborative, hatami2023status, stamatakis2021autonomous, liu2023optimizing, leng2019age, gong2020age} for pull-based schemes. In \cite{shao2021partially, tahir2024collaborative, gong2020age, liu2023optimizing}, system models involving multiple sensors and a single monitor have been studied for various scenarios. In these works, the monitor is only aware of the AoI of the sensor that successfully transmits at that time slot and estimates the AoI of other sensors based on its previous observations. In \cite{shao2021partially}, \cite{gong2020age} and \cite{liu2023optimizing}, sub-optimal policies are obtained via the index policy, the myopic policy, and the particle filter method, respectively. A similar system model has been studied in \cite{chen2022uncertainty} in which an entropy-based metric, namely \emph{uncertainty of information}, is minimized by employing an index policy. Additionally, AoI minimization problems in \cite{stamatakis2021autonomous, leng2019age, hatami2023status} consider failure status, channel availability, and the battery level as unobservable states, respectively. As a final related work, the authors of \cite{zhao2025optimizing} and \cite{chiariotti2025goal} consider a push-based scheme where sensors estimate the freshness level of others to decide when to transmit. 

\section{Preliminaries}\label{sec:prelim}
\subsection{Semi-Markov Decision Process (SMDP)}\label{sec:SMDP}
This section describes the SMDP framework and presents the policy iteration method we utilize in this paper, which is based on \cite{book_ross} and \cite{book_ibe}. The discrete-time SMDP of interest to the current paper is the 5-tuple $(\mathcal{S},\mathcal{A},Q,C,H)$ described below.
\begin{itemize}
    \item The {\em state space} $\mathcal{S} = \{ 1,2,\ldots,S \}$ of the SMDP is the finite set of states that are visited by the SMDP process.
    \item The {\em action space} $\mathcal{A} = \{ 1, 2, \ldots, A \}$ is the finite set of $A$ actions that can be taken at a state. 
    The use of different action sets at different states is not of interest to this paper, and is therefore not discussed.
    \item $Q: \mathcal{S} \times \mathcal{A} \times \mathcal{S} \rightarrow [0, 1]$  is called the {\em transition function} of the SMDP. Particularly, when action $a \in A$ is taken at current state $s$, a transition to state $s'$ occurs with probability $Q(s,a,s')$.
    \item $C: \mathcal{S} \times \mathcal{A} \rightarrow [0, \infty)$ is the {\em cost function} which represents the average accumulated cost $C(s,a)$ when action $a$ is taken at state $s$, until the transition to the next state.
    \item $H: \mathcal{S} \times \mathcal{A} \rightarrow (0, \infty)$ is the {\em holding time function} 
    which is the average time spent at state $s$ when action $a$ is taken, which is denoted by $H(s,a)$.
\end{itemize}
A deterministic policy $\pi : S \rightarrow A$ is one that maps each state $s \in S$ to a single action $a \in A$, i.e., we take the action $\pi_s$ when we are at state $s$. Let a deterministic policy $\pi$ be given. Let the embedded (at the decision epochs) Markov chain associated with the policy $\pi$  be denoted by $Y_t^\pi, t \geq 0$, where $Y_t^{\pi}$ is the state of the system at the $t$th decision epoch. Let $Z_t^\pi$ denote the total cost accumulated up to time $t, t \geq 0$. If $Y_t^\pi$ has no two disjoint communicating classes, then for each initial state $Y_0^\pi=s$, the limit 
\begin{align}
    \lim_{t \rightarrow \infty} \frac{Z^{\pi}(t)}{t} = C^{\pi}, \label{smdp}
    \end{align}
exists and is independent of the initial state $s$. Another interpretation for $C^{\pi}$ is through the steady-state probabilities $x_s^{\pi} = \lim_{t \rightarrow \infty} \mathbb{P} (Y_t^{\pi} =s)$, $s=1,\ldots,S$,
\begin{align}
    C^{\pi} & = \frac{\sum_{s=1}^S x_s^{\pi} C(s,\pi_s)}{\sum_{s=1}^S x_s^{\pi} H(s,\pi_s)}.
\end{align}
Hence, the quantity $C^{\pi}$ amounts to the long-run average cost per unit time when policy $\pi$ is followed for the underlying SMDP. 

The deterministic policy $\pi^*$ that minimizes the long-run average cost in \eqref{smdp} is the optimum solution for the SMDP problem. Algorithm~\ref{alg:SMDP} presents the pseudo-code for the policy iteration algorithm to obtain $\pi^*$  given the SMDP described by the tuple $(\mathcal{S},\mathcal{A},Q,C,H)$ \cite{book_ross,book_tijms,book_ibe}.

\begin{algorithm}[tb]
    \caption{Policy Iteration Algorithm for an SMDP}
    \begin{algorithmic}
    \renewcommand{\algorithmicrequire}{\textbf{Input:}}
    \renewcommand{\algorithmicensure}{\textbf{Output:}}
        \REQUIRE  SMDP described by the tuple $(\mathcal{S},\mathcal{A},Q,C,H)$
        \ENSURE Policy $\pi^*$
        \STATE {\bf Step 1:} Start with an initial arbitrary policy $\pi$
         \STATE{\bf Step 2: (Relative Value Determination)} For the present policy $\pi$, solve the following linear system of equations for $s=1,2,\ldots,S$, for the relative values $V_s$ and the long-run average cost $C^{\pi}$ of the policy $\pi$,
        \begin{align}
          C^{\pi} H(s,\pi_s) + V_s = C(s,\pi_s) + \sum_{s'=1}^S Q(s,\pi_s,s') V_{s'},    \label{eq:poly_it}
        \end{align}
         by setting $V_S=0$
         \STATE{\bf Step 3: (Policy Improvement)} For each state $s=1,2,\ldots,S$, set the alternative action $a_s$ to,
        \[
        \underset{a}{\text{arg min}} \ \ {\frac{1}{H(s,a)} \left( C(s,a) + \sum_{s'=1}^S Q(s,a,s') V_{s'} - V_s \right)}   
        \]
        \STATE {\bf Step 4:} Update the policy $\pi_s=a_s$ for $s=1,2,\ldots,S$
        \STATE {\bf Step 5:} Stop when the two successive policies are the same and set $\pi^*=\pi$. Otherwise, go to Step 2.   
    \end{algorithmic}
    \label{alg:SMDP}
\end{algorithm}

\subsection{Absorbing Markov Chains and Time to Absorption}
The following is based on \cite{book_kemeny} and \cite{telek_book}. Consider a discrete-time Markov chain (DTMC) $X_t, t \geq 0$, with $U$ transient states and $V$ absorbing states with a probability transition matrix of the form
\begin{align}
    \begin{bmatrix}
    A & B \\
    0 & 0
    \end{bmatrix}, \label{AMC}
\end{align}
where $A$ is a square matrix of size $U$, and $B$ is a $U \times V$ matrix, comprising the transition probabilities among the transient states, and from the transient states to the absorbing states, respectively. From any transient state, the probability of eventually entering an absorbing state is one. Therefore, the fundamental matrix $(I-A)^{-1}$ exists where $I$ denotes the identity matrix. The DTMC $X_t$ has the initial probability vector of size $1 \times U$ denoted by ${\beta}$ whose $i{\text{th}}$ element is defined as, 
\begin{align}
    \beta_i & = \mathbb{P} (X_0 = i), i=1,\ldots,U. 
\end{align}
The probability that absorption occurs at absorbing state $v$, $v=1,2,\ldots,V$, denoted by $p_v$, is given by \cite{book_kemeny},
\begin{align}
  p_v & = \beta (I-A)^{-1}B(:,v),  
\end{align}
where $B(:,v)$ is the $v{\text{th}}$ column of $V$. We are also interested in the probability mass function (pmf) of the absorption time $W$ of this AMC,
$p_W(n)=\mathbb{P} (W =n), n\geq 1$, is given by the following expression, \cite{telek_book,alfa_book},
\begin{align}
    p_W(n) & = {\beta} {A}^{n-1} {b},  \label{phase_type}
\end{align}
where $b=Be$ and $e$ denotes a column vector of ones of suitable size.
The random variable $W$ with pmf given in \eqref{phase_type} is said to have a discrete phase-type (DPH) distribution.
Moreover, the first two factorial moments of $W$ can be written in closed-form through the following expressions \cite{telek_book}:
\begin{align}
    \mathbb{E}[W] & = {\beta} ({I}-{A})^{-1} {e},  \label{meanW} \\ 
    \mathbb{E}[W(W-1)] & = 2 {\beta} ({I}-{A})^{-2} {A} {e}. \label{secondW} 
\end{align}
\section{System Model} \label{sec:system}
We consider the time-slotted remote estimation system involving an information source and a monitor in Fig.~\ref{fig:sys}. The source observes a random process represented by a finite-state irreducible discrete-time Markov chain with $N$ states, denoted by $X_t \in \{1,2,\ldots,N\} $, with the transitions among the states governed by the transition matrix ${P}$. The dynamics of the source process are known to the monitor. We consider a pull-based scheme for which the monitor takes an action $a_t$ at the start of time slot $t$ to send a pull request, i.e., $a_t=1$, or stays idle, i.e., $a_t=0$, where pull requests are assumed to arrive at the source instantaneously. 

For the ZD channel, the monitor obtains the observation $o_t \in \mathcal{O}(a_{t})$ from the observation space $\mathcal{O}(0)=0$, i.e., no observation, and $\mathcal{O}(1) \in \{1,2,\dots,N\}$ as,
\begin{align}
    o_t=\begin{cases}
        X_{t}, & a_{t}=1, \\
        0, & a_{t}=0.
    \end{cases} \label{eq:obs1}
\end{align}
The occurrence order of the events when $X_{t-1}=j$ is given below.
\begin{enumerate}
    \item We move on to the next slot, i.e., time slot $t$.
    \item The state $j$ of the Markov chain is updated to $j'$ according to $P$.
    \item If $a_{t}=1$, the state $j'$ is transmitted and is immediately received by the monitor. Otherwise, there are no transmissions.
    \item The monitor updates its estimation $\hat{X}_t$.
\end{enumerate}

On the other hand, in the UD model, the channel introduces a one-slot delay, or equivalently, it takes one slot to transmit the information packet. 
In this case,
\begin{align}
    o_t=\begin{cases}
        X_{t-1}, & a_{t-1}=1, \\
        0 , & a_{t-1}=0.
    \end{cases} \label{eq:obs}
\end{align}
The way events occur when $X_t=j$ is according to the following order. 
\begin{enumerate}
    \item If $a_t=1$, the observation $j$ is transmitted. Otherwise, there are no transmissions.
    \item The state $j$ of the Markov chain $X_t$ is updated to $j'$ according to $P$.
    \item We wait for one slot to complete the transmission by moving on to the next slot $t+1$, and the monitor receives the new observation if any.
    \item The monitor updates $\hat{X}_{t+1}$.
\end{enumerate}

For both ZD and UD models, the estimation rule is the following. By using all past observations, the monitor obtains the distribution
\begin{align}
    \bm{\pi}_t & = \begin{bmatrix} \pi_t(1) & \pi_t(2) & \cdots & \pi_t(N) \end{bmatrix}, \label{eq:pi_t}
\end{align} 
where $\pi_t(n)=\mathbb{P}(X_t=n \, | \, \mathcal{H}_t)$, and
\begin{align}
    \mathcal{H}_t=\{o_1,\dots,o_t\}, \label{eq:hist}
\end{align}
represents all observations made until time $t$. The monitor either employs the MAP estimator $\hat{X}_t$, 
\begin{align}
    \hat{X}_t=\arg\max \bm \pi_t,  \label{MAP}
\end{align}
which is the most likely state for the original process $X_t$ from the perspective of the monitor, or the martingale estimator, which is the most recently received observation. Again, for both models, immediately after the monitor updates its estimate, the AoII process of the remote estimation system at time $t$, denoted by $\text{AoII}_t$, is also updated. Particularly, $\text{AoII}_t$ penalizes the estimation error linearly while the error stays, and it is reset to zero when $X_t$ and $\hat{X}_t$ are synchronized, whose evolution can then be expressed mathematically as,
\begin{align}
    \text{AoII}_{t+1}=\begin{cases}
        \text{AoII}_{t}+1, & X_t\neq \hat{X}_t, \\
        0, & X_t=\hat{X}_t.
    \end{cases} \label{eq:aoii}
\end{align}
In this paper, we refer to \eqref{eq:aoii} as the evolution equation for the AoII process, but more general evolution equations under the unifying AoII framework of \cite{maatouk2020} are not considered in this paper. Notice that the AoII update is a system-wide update, and cannot be carried out by the monitor alone. The reason for this is that the monitor is not AoII-aware with the exception of the pull instants in the ZD model when the AoII process is reset to zero, which is known to the monitor. Let AoII be the steady-state random variable for the asymptotically stationary random process $\text{AoII}_t$, i.e.,
\begin{align}
    \mathbb{P} (\text{AoII} \leq x) = \lim_{t \rightarrow \infty} \mathbb{P}(\text{AoII}_t \leq x). 
\end{align}
The mismatch between $X_t$ and $\hat{X}_t$ is quantified by using the mean of AoII, denoted by MAoII,
\begin{align}
    \text{MAoII}= \mathbb{E} [\text{AoII}] = \lim_{T\to\infty} \frac{1}{T}\sum_{t=1}^T\text{AoII}_t,
\end{align}
where the last equality stems from the ergodicity of the AoII process. The long-run sampling rate is denoted by $R$,
\begin{align}
    R & =\lim_{T\to\infty} \frac{1}{T} \sum_{t=0}^{T-1} a_t.   
\end{align}
We consider a sampling constraint on the monitor such that it should not send pull requests every time slot due to energy consumption considerations. With a budget $\alpha$ for the average sampling rate, the constrained optimization problem is cast as,
\begin{mini}
	{\phi}{\text{MAoII}^{\phi}} 
	{\label{Opt1}}
    {}
	\addConstraint{ R^{\phi} }{\leq \alpha,} 
\end{mini}
where $\text{MAoII}^{\phi}$ and $R^{\phi}$ denote the mean AoII and sampling rate obtained when the policy $\phi$ is imposed over a certain set of policies, which may be problem-dependent. This optimization problem can be converted to an unconstrained problem with a Lagrangian coefficient $\lambda$ as,
\begin{mini}
	{\phi}{\text{MAoII}^{\phi}+\lambda R^{\phi},} 
	{\label{OptUC}}
    {}
\end{mini}
and the optimum solution to the unconstrained problem \eqref{OptUC} for a specific value of $\lambda$, denoted by $\lambda^*$, is a solution to the constrained problem \eqref{Opt1}. We note that bisection search can be used to find $\lambda^*$ to solve the constrained SMDP problem \eqref{Opt1} \cite{makowski1986,cosandal_etal_TRIT24}.

\section{Zero Delay Channel}\label{sec:zerodelay}
\begin{figure}[t]
    \centering
    \begin{tikzpicture}[x=1.15cm, y=0.85cm, font=\small, scale=0.8]
    
    
    \draw[->, thick] (-0.1,0) -- (8.7,0) node[right] {$t$};
    \draw[->, thick] (0,-0.2) -- (0,4.5) node[above] {$\mathrm{AoII}_t$};
    
    \draw[aoii line] (0,0) -- (2,0);   
    \draw[aoii line] (2,0) -- (5,3);   
    \draw[aoii line] (5,3) -- (6,0);   
    \draw[aoii line] (6,0) -- (8,2);   
    
    \node[dot] at (0,0) {};
    \node[dot] at (1,0) {};
    \node[dot] at (2,0) {};
    \draw[drop](3,1)--(3,0); \node[dot] at (3,1) {};
    \draw[drop](4,2)--(4,0); \node[dot] at (4,2) {};
    \draw[drop](5,3)--(5,0); \node[dot] at (5,3) {};
    \node[dot] at (6,0) {};
    \draw[drop](7,1)--(7,0); \node[dot] at (7,1) {};
    \draw[drop](8,2)--(8,0); \node[dot] at (8,2) {};
    
    \foreach \y in {1,2,3}{
    \draw[thin](-0.07,\y)--(0.07,\y);
    \node[left,font=\scriptsize] at (-0.12,\y) {\y};
    }
    \node[left,font=\scriptsize] at (-0.12,0) {$0$};
    
    \foreach \t in {0,1,...,8}{
    \draw[thin](\t,0.07)--(\t,-0.07);
    \node[below,font=\scriptsize] at (\t,-0.10) {$\t$};
    }
    
    \foreach \x/\xl in {
      0/{$(1,0)$}, 1/{$(1,0)$}, 2/{$(1,0)$},
      3/{$(2,1)$}, 4/{$(3,2)$}, 5/{$(3,3)$},
      6/{$(1,0)$}, 7/{$(3,1)$}, 8/{$(3,2)$}}{
    \node[font=\scriptsize] at (\x,-1.00) {\xl};
    }
    \end{tikzpicture}
    \caption{An example sample SMDP cycle for the process $(X_t,\mathrm{AoII}_t)$ for $k=9$. The SMDP cycle starts at $t=0$ with an initial pull giving $\hat{X}_0=X_0=1$ and $\mathrm{AoII}_0=0$. The martingale estimator keeps $\hat{X}_t=1$ throughout the cycle $t=0,\ldots,k{-}1=8$. At $t=6$, the source returns to state $1$, causing a spontaneous reset $\mathrm{AoII}_6=0$ without any transmission. The next pull occurs at $t=k=9$, ending the cycle. The values of $(X_t,\mathrm{AoII}_t)$ are given below the $t$-axis.}
    \label{fig:ZD_path}
\end{figure}
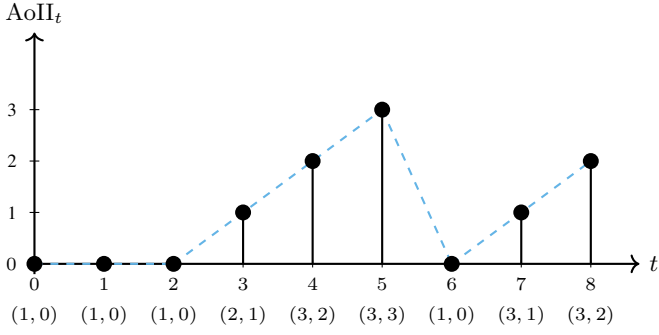

Revisiting the ZD transmission model, consider an observation-dependent policy for which we sample the original process $X_t=i$ at some arbitrary time, say at time $0$, which leads to $\hat{X}_0=i$ and $\text{AoII}_0=0$ since the source and monitor are in sync at time $0$. Let us now assume that the next sample is to be taken in $k \leq K$ slots, where $K$ is the maximum allowed inter-sampling time, and the wait $k$ should be dependent on the observation $i$, which makes the studied policy dependent on observations. Our goal is to develop an SMDP model represented by the tuple $(\mathcal{S},\mathcal{K},Q,C,H)$ which is now described. Here, $\mathcal{S} = \{ 1,2,\ldots,N \}$ is the same state space of the original process, and $\mathcal{K} = \{ 1, 2, \ldots, K \}$ is the finite set of $K$ actions $k \in \mathcal{K}$ that can be taken at a state. The transition function $Q$ of the underlying SMDP is written as,
\begin{align}
    Q(i,k,j) & = \mathbb{P} (X_{k}=j \, | \, X_0=i)\\
    & = e_i^{T} P^k e_j,
\end{align} 
where $e_j$ is a column vector of zeros except for the $j{\text{th}}$ entry, which is one, whereas the holding time function $H(i,k)=k$ amounts to the inter-sampling time, which is the action itself. On the other hand, the cost function $C(i,k)$ is more involved to write. The martingale estimator is given by $\hat{X}_t=i$ from the time the last sample was taken to the next sample, whereas the MAP estimator is written as $\hat{X}_t= \text{arg max } e_i^{T} P^t$. Let us now consider the following two-dimensional Markov process $\{ (X_t,\text{AoII}_t), \ 0 \leq t < k, \ 0 \leq \text{AoII}_t < k \}$. An example sample path of how this process evolves in time is illustrated in Fig.~\ref{fig:ZD_path}. Initially, $X_0=i$ and $\text{AoII}_0=0$. A new sample is taken when $t=k$ upon which $\text{AoII}_k=0$. Let us assume that at time $t<k$ we are at state $(j,\delta)$. We transition to state $(j',\delta+1)$ with probability $q_{j,j'}$ if $j' \neq \hat{X}(t)$ and to state $(j',0)$ with probability $q_{j,j'}$ if $j' = \hat{X}(t)$. Having constructed the probability transition matrices of the non-homogeneous Markov chain for each step $t, \ 1 \leq t < k$, one can obtain the probability $\mathbb{P}(X_t=j,\text{AoII}_t=\delta), 1 \leq j \leq N, 0 \leq \delta < k$ using the transient solution of the underlying chain with known initial probabilities. Consequently, we write the cost function $C(i,k)$ as
\begin{align}
    C(i,k) = \lambda + \sum_{t=1}^{k-1} \sum_{j=1}^{N} \sum_{\delta=1}^{k-1} \delta \ \mathbb{P}(X_t=j,\text{AoII}_t=\delta).  \label{eq:Cost}
\end{align}
The first term stems from the observation that we take one single sample during an inter-sampling period, i.e., an SMDP cycle. The second term is the expected value of the total AoII accumulated in the same period. At this point, we have completed the SMDP formulation for which we propose to employ policy iterations given in Algorithm~\ref{alg:SMDP} to obtain the optimum deterministic policy $\pi^*$ with cost $C^{\pi^*}$ that minimizes the cost function \eqref{OptUC} for a given value of the Lagrangian parameter $\lambda$. Then, we use bisection search to find the particular value of $\lambda=\lambda^*$ that yields the optimum solution for the constrained SMDP \eqref{Opt1}. 

\section{Unit Delay Channel} \label{sec:unitdelay}

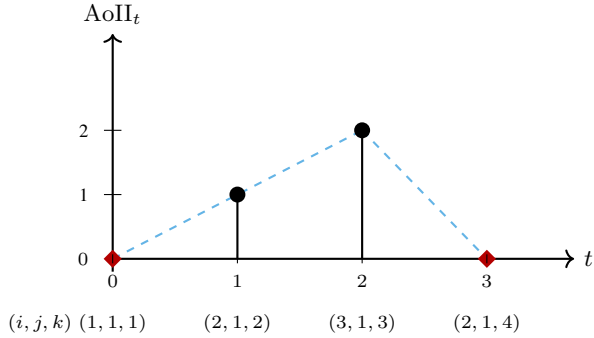
\begin{figure}[t]
    \centering
    \begin{tikzpicture}[x=1.65cm, y=0.85cm, font=\small]
    
    \draw[->, thick] (-0.05,0) -- (3.7,0) node[right] {$t$};
    \draw[->, thick] (0,-0.2) -- (0,3.5) node[above] {$\mathrm{AoII}_t$};
    
    \draw[aoii line] (0,0) -- (2,2);   
    \draw[aoii line] (2,2) -- (3,0);   
    
    \node[reddia] at (0,0) {};
    
    \draw[drop](1,1)--(1,0); \node[dot] at (1,1) {};
    \draw[drop](2,2)--(2,0); \node[dot] at (2,2) {};
    
    \node[reddia] at (3,0) {};
    
    \foreach \y in {1,2}{
    \draw[thin](-0.07,\y)--(0.07,\y);
    \node[left,font=\scriptsize] at (-0.12,\y) {\y};
    }
    \node[left,font=\scriptsize] at (-0.12,0) {$0$};
    
    \foreach \t in {0,1,...,3}{
    \draw[thin](\t,0.07)--(\t,-0.07);
    \node[below,font=\scriptsize] at (\t,-0.10) {$\t$};
    }
    
    \node[left,font=\scriptsize] at (-0.25,-1.00) {$(i,j,k)$};
    \foreach \x/\xl in {
      0/{$(1,1,1)$},
      1/{$(2,1,2)$},
      2/{$(3,1,3)$},
      3/{$(2,1,4)$}}{
    \node[font=\scriptsize] at (\x,-1.00) {\xl};
    }  
    \end{tikzpicture}
    \caption{An example path for the process $Z_t=(i,j,k)$ with $g(1)=5$ and the MAP estimator $f(1,k)=1$ for $k\leq 3$ and $f(1,k)=2$ for $k>3$. The cycle starts at $t=0$ from the synchronization state $(1,1,1)$ since $X_0=1=f(1,1)=1$. At $t=1$ and $t=2$, the states $(3,1,2)$ and $(3,1,3)$ are transient since $X_t\neq f(1,k)$, causing AoII to increase by one each slot. At $t=3$, the source transitions to state $2$ and the estimate switches to $f(1,4)=2$, synchronizing the monitor. Hence, the AMC is absorbed into the synchronization state $(2,1,4)$. The sync states are marked with red diamonds and $(i,j,k)$ values are depicted below the $t$-axis.}
    \label{fig:UD_path}
\end{figure}

In this section, we first analytically model the remote estimation system when an observation-dependent pull policy is deployed. The observation-dependent policy to be studied is the one for which we wait for $g(j)$ slots since the last pull request resulted in observation $j$. We also let $K= \max_j g(j)$. This analytical model enables us to calculate the MAoII and average sampling rate $R$ for given benchmark policies that are dependent on observations. Subsequently, we develop a belief-dependent policy using CDRL, which can be studied by simulations.

\subsection{Analytical Model for Observation-Dependent Pull Policies} \label{sec:ODpull}
In this subsection, we will model the remote estimation system within the embedded epochs of synchronization, namely a single AoII cycle, through an absorbing Markov chain (AMC) $\{ Z_t = (i,j,k), t \geq 0, 1 \leq i,j \leq N, 1 \leq k \leq K \}$ with multiple absorbing states. Here, $i$ denotes the state of the source, $j$ denotes the last observation made, and $k$ denotes the time since the pull request resulting in observation $j$. Fig.~\ref{fig:UD_path} illustrates an example path for the described process $Z_t$. The MAP estimator is the same as the previous subsection and can be written as a function of $j$ and $k$, denoted by $f(j,k) =  \text{arg max } e_j^{T} P^k$, whereas for the martingale estimator, $f(j,k)= j$, for $1 \leq k \leq K$. The states $(i,j,k)$ satisfying $i=f(j,k)$ are the synchronization states (or absorbing states), and the others, when $i \neq f(j,k)$  are the transient states. Let us assume that we have $L$ synchronization states and therefore we have $V = S-L$ transient states, where $S=K N^2$. The AMC starts operation at $t=0$ from a synchronization state. In one slot, the age can remain at zero, in which case a new AoII cycle will be initiated without visiting a transient state, or the age will start to increase, visiting the transient states until the source and monitor are synchronized at some time point, upon which the AMC is to be absorbed into one of the synchronization states and a new AoII cycle will be initiated.   

The transition probabilities of the AMC are now given. Let us assume that the AMC is initially in a synchronization state $l \in \{ 1,2,\ldots,L \}$ where the synchronization states are obtained by suitable indexing (or ordering) of the triple $(i,j,k)$ satisfying $i=f(j,k)$. During the evolution of the AMC, let us assume that we are at an arbitrary transient state $(i,j,k)$. Then, the following transitions are possible:
\begin{itemize}
    \item When $k \geq g(j) $, we transition to transient state $(i',i,1)$ with probability $q_{i,i'}$ if $f(i,1) \neq i'$, or to synchronization state $({i'},{i},1)$ with probability $q_{i,i'}$ if $f(i,1) = i'$.  The former transition indicates a situation when an observation is received but synchronization cannot be achieved. However,  in the latter transition, synchronization is achieved thanks to the pull request.
    \item  When $k < g(j)$, we transition to synchronization state $(i',j,k+1)$ with probability $q_{i,i'}$ if $f(j,k) = i'$, or to transient state $({i'},{j},{k+1})$ with probability $q_{i,i'}$ if $f(j,k) \neq i'$.  The former transition indicates a situation when an observation is not received and synchronization does not take place on its own. In the latter transition, synchronization is achieved without the support of a pull request.
\end{itemize}
With proper ordering of the joint states, the probability transition matrix of the AMC is in the same form as \eqref{AMC} where $A$ and $B$ are square matrices of size $S$. Let $W^{(l)}$ be the time to absorption when the AMC starts from the synchronization state $l$ which corresponds to a probability vector $\beta^{(l)}$ which is a row vector of zeros of size $S$ except for one single entry which is one, corresponding to a particular value of the triple $(i,j,k)$ such that $i=f(j,k)$. The total accumulated expected age in this cycle, denoted by $C(l)$, is given by \eqref{secondW},
\begin{align}
    C(l) & = \mathbb{E}[1+2+ \cdots + (W^{(l)}-2)+ (W^{(l)}-1)] \\ 
    & = \mathbb{E} [W^{(l)}(W^{(l)}-1)/2] \\
    & = \beta^{(l)} (I-A)^{-2} Ae. \label{eq:cost_age}
\end{align}
On the other hand, the holding time at state $l$, denoted by $H(l)$, is written as,
\begin{align}
    H(l) & = \mathbb{E} [W^{(l)}] = \beta^{(l)} (I-A)^{-1} e. 
\end{align}
Let $l'$ be given as an arbitrary synchronization state. Let us denote the column of $B$ corresponding to the synchronization state $l'$ by $B^{(l')}$. In this case, the probability of transitioning from synchronization state $l$ to $l'$ is given by $Q(l,l')$,
\begin{align}
    Q(l,l') & =  \beta^{(l)} (I-A)^{-1} B^{(l')}.
\end{align}
The above formulation constructs a semi-Markov process (SMP) which is a sub-case of the SMDP formulation described in Subsection~\ref{sec:SMDP} without actions, or with a single deterministic policy. Therefore, we adopt \eqref{eq:poly_it} from the policy evaluation step of Algorithm~\ref{alg:SMDP} to find the average cost for this problem, which in turn gives us the MAoII and the average sampling rate of the remote estimation system, based on our cost definition. Particularly, if we solve the following $L$ linear equations, for $l \in \{ 1,2,\ldots,L \}$,
\begin{align}
    \hat{C} H(l) + V_l = C(l) + \sum_{l'=1}^L Q(l,l') V_{l'},
\end{align}
with $V_L=0$, $\hat{C}$ gives the MAoII when $C(l)$ is written as in \eqref{eq:cost_age}, and the sampling rate when we take $C(l)=1$.

\subsection{Belief-based CDRL} \label{sec:CDRL}
Since the monitor will never have the actual state of the source at any time due to the communication delay, it also would not know the instantaneous value of AoII$_t$ with certainty.  A successful transmission cannot even guarantee a reset of the AoII for the UD channel since the source process may have changed and consequently the information becomes stale before the monitor updates its estimation. However, the monitor can estimate the joint distribution of AoII$_t$ and the state, which is termed as \emph{belief}, based on the observations it receives. In this section, we first construct and then exploit the belief to formulate the optimization problem as a belief-MDP. First, we define the belief for this problem and explain how it evolves along with the observations. Then, the belief-MDP formulation is developed using the transition probabilities among beliefs under an action, and the expected cost of a belief-action pair. Finally, we propose a CDRL algorithm to solve the belief-MDP.

\subsubsection{Belief Evolution} \label{sec:BUpd} 
We denote the belief at time $t$ by $b_t$, and it can be expressed as 
\begin{align}
    b_t(j,\delta)=\mathbb{P}(X_t=j, \text{AoII}_t=\delta | \mathcal{H}_t), \label{eq:b}
\end{align}
where $\mathcal{H}_t$ is the collection of all observations until time $t$, and is defined in \eqref{eq:hist}. The monitor calculates this belief in two steps as described below. In the first step, consider that the monitor has the belief $b_{t-1}$ at the beginning of time slot $t$, and a new observation $o_t$ arrives. Notice from \eqref{eq:obs} that if $a_{t-1}=1$, the new observation includes information from $X(t-1)$, thus the belief should be updated accordingly. We define the updated belief as $\hat{b}_t(j,\Delta)=\mathbb{P}(X_t=j, \text{AoII}_t=\Delta|\mathcal{H}_{t+1})$ and the update can be performed as, 
\begin{align}
    \hat{b}_{t}(j,\delta)&=\mathbb{P}(X_{t}=j, \text{AoII}_{t}=\delta|\mathcal{H}_{t+1}) \\
    &=\dfrac{ b_{t}(j,\delta)}{\sum_{d=0}^{\Delta}\sum_{k=1}^N  b_{t}(k,d)}. \label{eq:hat_b_t} 
\end{align}
Here, for practical reasons, we truncate the AoII values to $\Delta$, and assume that the AoII value does not further increase upon reaching $\Delta$. Notice that this truncation step was not needed in the SMDP formulation for the ZD channel.

\begin{figure}[t]
    \centering
    \vspace{0.1in}
    \includegraphics[width=0.8\linewidth]{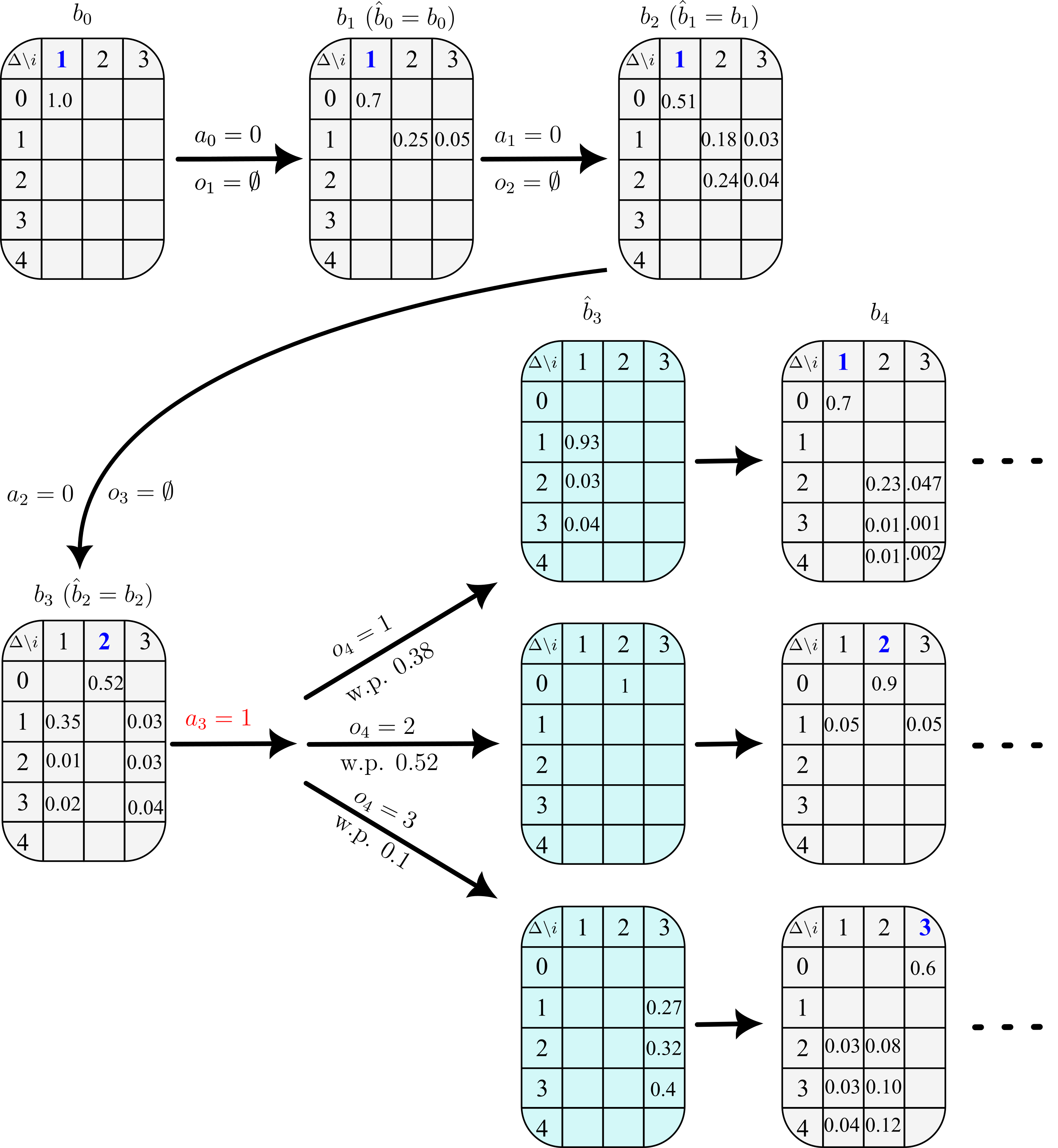}
     \caption{Evolution of the belief for a ternary source whose probability transition matrix is given in \eqref{eq:P_fig}. The process starts from state $1$. At each time slot, the monitor estimates the state with the highest likelihood, which is highlighted in blue. For the first two time slots, the monitor stays idle, and at the third time slot, the monitor sends a pull request. After receiving an observation at time slot $4$, the monitor first updates $\hat{b}_3$, and subsequently obtains the belief $b_4$ by using the update. All possible observations are illustrated with their relevant probabilities.}
    \label{fig:table}
\end{figure}

As \eqref{eq:aoii} indicates, if a state transition occurs towards the estimated state, AoII resets to zero. Otherwise, the AoII process is incremented by one. Consequently, the belief at time $t+1$ can be calculated using the updated belief $\hat{b}_t$, and the source dynamics as follows,
\begin{align}
    &b_{t+1}(j,\delta)=
    \begin{cases}
            \max  \bm{\pi}_{t+1}, &  j=\hat{X}_{t+1}, \delta=0, \\
           \begin{aligned} \sum_{k=1}^{N}\hat{b}_{t}(k,\delta\!-\!1)p_{kj},\end{aligned} &  j\neq\hat{X}_{t+1}, 0 < \delta < \Delta,\\
            \begin{aligned}
                &\sum_{k=1}^{N}\hat{b}_{t}(k,\delta\!-\!1)p_{kj} \\
                &+\sum_{j=1}^{N}\hat{b}_{t}(k,\delta)p_{kj},
            \end{aligned}
             &  j\neq\hat{X}_{t+1},  \delta=\Delta,\\
            0, & \text{otherwise.}
        \end{cases} \!\label{eq:b_t}
\end{align}
Notice that when the estimation is correct at time $t+1$, i.e., $X_{t+1}=\hat{X}_{t+1}$, $\text{AoII}_{t+1}$ becomes zero regardless of its previous value, and its probability can be obtained from \eqref{eq:pi_t} and source dynamics. This two-step process of constructing the belief is exemplified in Fig.~\ref{fig:table} for a process with transition matrix,
\begin{align}
    \begin{bmatrix} 
    0.70 & 0.25 & 0.05 \\
    0.05 & 0.90 & 0.05 \\
    0.10 & 0.30 & 0.60
    \end{bmatrix}. \label{eq:P_fig}
\end{align}

It is well-known that the belief $b_t$ contains all the necessary information about the entire observation history $\mathcal{H}_t$ regarding decision making, and thus, it is a sufficient statistic \cite{smallwood1973optimal}. Moreover, the next belief $b_{t+1}$ conditioned on the entire history of observations until time $t+1$ can exactly be obtained from the previous belief $b_t$ along with the new observation $o_t$ using \eqref{eq:hat_b_t} and \eqref{eq:b_t}. Therefore, the current belief and the next observation (outcome of the current action) can be mapped to the next belief with a function $\xi$ as $b_{t+1}=\xi(b_t,o_{t+1})$. In the example illustrated in Fig.~\ref{fig:table}, $b_2$ is obtained from $b_1$ and the observation $o_2=0$, thus $b_2=\xi(b_1,0)$. 

\subsubsection{Belief-MDP Formulation} 
For the belief-MDP formulation, we first define $T(b_{t+1},a_t,b_t)$ as the transition probability from $b_t$ to $b_{t+1}$ for a given action $a_t$. Although the belief space includes continuous-valued elements, the number of possible next beliefs is limited by the number of possible observations. In other words, for any belief $b_t$ and action $a_t$, there are $|\mathcal{O}(a_t)|$ possible next beliefs $b_{t+1}$. Additionally, the transition probability from belief $b_t$ to belief $b_{t+1}=\xi(b_{t},o_{t+1})$ for an action $a_t$ is equal to the probability of the observation $o_{t+1}$ for given $b_t$ and $a_t$, since $\xi(b_{t},o_{t+1})$ is a deterministic mapping. This can be expressed as
\begin{align}
    T(\xi(b_{t},o_{t+1}),b_t,a_t)=&\mathbb{P}(b_{t+1}=\xi(b_{t},o_{t+1})|b_t,a_t)\\=&\mathbb{P}(o_{t+1}|b_t,a_t). \label{eq:Tp}
\end{align}

It is worth reminding that after the action $a_{t}=0$ is taken, the observation $o_{t+1}=0$ occurs with probability $1$. Thus, when $a_t=0$, \eqref{eq:Tp} becomes
\begin{align}
    T(b_{t+1},b_t,0)=\begin{cases}1, &
        b_{t+1}=\xi(b_{t},0), \\
        0, & \text{otherwise.}
    \end{cases} \label{eq:tprob0}
\end{align}
When $a_{t}=1$, the probability of the observation $o_{t+1}=j, j\in\{1,\dots,N\}$ equals the probability that the process is at state $j$ at time $t$, which has already been defined by $\pi_t(j)$. Thus, we express the transition probability for $a_t=1$ as 
\begin{align}
    T(b_{t+1},b_t,1)=\begin{cases}
        \pi_t(j), & b_{t+1}=\xi(b_t,j), \\
        0, & \text{otherwise.}
    \end{cases} \label{eq:tprob1}
\end{align}

From \eqref{eq:Cost}, we define the expected cost of the belief state $b_t$ and action $a_t$ for any $\lambda$ as the expected AoII at time $t$ plus the sampling cost if the action $a_t=1$, and it can be expressed as
\begin{align}
c_{\lambda}(b_t,a_t)&=\mathbb{E}[\mbox{AoII}_t]+\lambda a_t \\ &=\sum_{\delta=0}^{\Delta} \delta \sum_{j=1}^N\mathbb{P}(X_t=j,\mbox{AoII}_t=\delta|\mathcal{H}_t)+\lambda a_t\\
    &=\sum_{\delta=0}^{\Delta}\delta\sum_{j=1}^N b_t(j,\delta)+\lambda a_t. \label{eq:cost}
\end{align}

Now, we can summarize the belief-MDP by using formulations in \cite{dagan2024resolving, lim2023optimality} as a tuple $\left(\mathcal{B},\mathcal{A},T,c,\gamma\right)$:
\begin{itemize}
    \item The belief space is $\mathcal{B}=[0,1]^{ N(\Delta+1)}$, and the belief $b_t \in \mathcal{B}$ is defined as in \eqref{eq:b}.
    \item The action space of the problem is defined as $\mathcal{A}=\{0,1\}$.
    \item After the reception of an observation, the state distribution is updated using \eqref{eq:pi_t} and the state transition matrix $P$, and the estimation is updated according to \eqref{MAP}.
    \item For each action $a_t$, there are $|\mathcal{O}(a_t)|$ possible observations, hence that many next beliefs with non-zero transition probabilities. Each observation $o_{t+1}$ forces the belief $b_t$ to change according to $b_{t+1}=\xi(b_t,o_{t+1})$ with probability $T(\xi(b_t,o_{t+1}),a_t,b_t)$. The derivation of $\xi(b_t,o_{t+1})$ is detailed in the steps \eqref{eq:hat_b_t} and \eqref{eq:b_t}, and the transition probabilities are derived in \eqref{eq:tprob0} and \eqref{eq:tprob1}.
    \item The cost of a belief-action pair $(b_t,a_t)$ for any $\lambda$ is defined through a cost function $c_{\lambda}(b_t,a_t)$ in \eqref{eq:cost}.
    \item The parameter $\gamma$ is the discount factor.
\end{itemize}

Finally, Bellman's optimality equation \cite{spaan2012partially} for the unconstrained problem in \eqref{OptUC} can be expressed as
\begin{align}
    c_{\lambda}(b_t,a_t) + \gamma\sum_{o_{t+1} \in \mathcal{O}(a_t)} T(\xi(b_t,o_{t+1}),a_t,b_t)  V(\xi(b_t,o_{t+1})), \label{eq:bell}
\end{align}
where $V(b_t,a_t)$ is the average cost attained when the action $a_t$ is applied first at the initial belief $b_t$, followed by the optimal policy thereafter, and $V(\xi(b_t,o_{t+1}))$ is the minimum of it over all actions, i.e., $V(\xi(b_t,o_{t+1}))=\min_{a'\in\{0,1\}}V(\xi(b_t,o_{t+1}),a')$.

\subsubsection{Constrained Deep Reinforcement Learning}
\begin{figure}[t]
    \centering
    \includegraphics[width=0.90\linewidth]{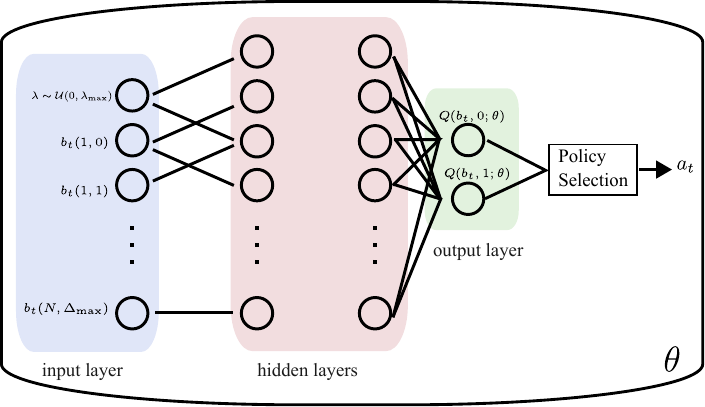}
    \caption{Network architecture used for DRL with parameter $\theta$.}
    \label{fig:RL}
\end{figure}

Since the belief space includes uncountable elements, we cannot obtain the value function for each belief with dynamic programming from the belief-MDP representation. Therefore, we use a non-linear approximation for the value function in \eqref{eq:bell}. The network architecture includes fully connected layers and a policy selection step as summarized in Fig.~\ref{fig:RL}. Details of the network and proposed algorithm are explained as follows.

\paragraph{State Augmenting}
To solve the constrained problem, we adapt the state-augmenting approach for CDRL from \cite{naderializadeh2022state}. We define the augmented state $z_t$ by augmenting the Lagrangian coefficient as $z_t=(b_t,\lambda)$. In the training phase, we feed the network with $\lambda$, which is a uniformly generated random variable from the interval $[0,\lambda_{\max}]$. In the execution phase, the dual variable is initiated as $\lambda=0$, and it is dynamically updated after each $B$ samples. The $k$th update of $\lambda$ can be expressed as
\begin{align}
    \lambda\to \max\left(0,\lambda + \eta_{\lambda} \left(\beta_k-\alpha\right)\right), \label{eq:lambda_upd}
\end{align}
where $\beta_k=\sum_{t=(k-1)B+1}^{kB} a_t$ is the sampling rate from the last $B$ samples, and $\eta_{\lambda}$ is the update rate of $\lambda$.

\paragraph{Target Network}
As suggested in \cite{egorov2015deep} and \cite{lillicrap2015continuous}, we construct two networks, namely, a main network and a target network, with parameters $\theta$ and $\theta'$, respectively, to improve the convergence of learning. The target network parameters $\theta'$ are updated with the main network parameters by the soft update rule proposed in \cite{lillicrap2015continuous}, which can be expressed as
\begin{align}
    \theta'\leftarrow \tau\theta+(1-\tau)\theta'.
\end{align}
The main network parameters $\theta$, on the other hand, are obtained by minimizing the cost function $J(\theta)$, which is written as
\begin{align}
        &\hat{V}(b_t,a_t;\theta)-\left(c_{\lambda}(b_t,a_t)+\gamma   \hat{V}_{next}(b_t,a_t;\theta')\right), \label{eq:costdqn}
\end{align}
where $\hat{V}_{next}(b_t,a_t:\theta')$ represents the average next state cost, obtained by the target network as
\begin{align}
    \sum_{o_{t+1} \in \mathcal{O}(a_t)} T(\xi(b_t,o_{t+1}),a_t,b_t) \hat{V}(\xi(b_t,o_{t+1});\theta').
\end{align}
Parallel to the notation in \eqref{eq:bell}, $\hat{V}(b_t,a_t;\theta)$ denotes the estimated cost for the belief $b_t$, and an action $a_t$ for the policy obtained from the main network $\theta$ is applied. Similarly, the expression $\hat{V}(\xi(b_t,o_{t+1});\theta')$ is obtained using the target network $\theta'$, i.e., $\hat{V}(\xi(b_t,o_{t+1});\theta')=\min_{a'\in\mathcal{A}}\hat{V}(\xi(b_t,o_{t+1}),a';\theta')$. The cost function is minimized by stochastic gradient descent \cite{amari1993backpropagation} with a learning rate $\eta$. 

\paragraph{Exploration Exploitation Tradeoff}
Throughout the learning phase, the policy is selected via exploration/exploitation with an exploration coefficient $0<\delta<1$, and a decaying coefficient $0<\nu\le 1$. In particular, with $e$ being the epoch number, the action minimizing $\hat{V}(b_t,a_t;\theta)$ is chosen with probability $(1-\delta) \nu^{e-1}$, and an action is chosen uniformly random with probability $\delta \nu^{e-1}$.

To summarize the algorithm, belief-CDRL consists of two phases i) training phase, ii) execution phase. In the training phase, both network parameters are trained for uniformly generated $\lambda$ values. In the execution phase, $\lambda$ values are updated by \eqref{eq:lambda_upd}, and the action
\begin{align}
    a_t= \underset{\{0,1\}}{\arg\min} \ c_{\lambda}(b_t,a_t)   +\gamma\hat{V}_{next}(b_t,a_t;\theta')
\end{align}
is applied.

\section{Numerical Results} \label{sec:Numerical}
\subsection{ZD Channel Examples}
\begin{figure}[t]
	\centering
    \includegraphics[width=0.8\columnwidth]{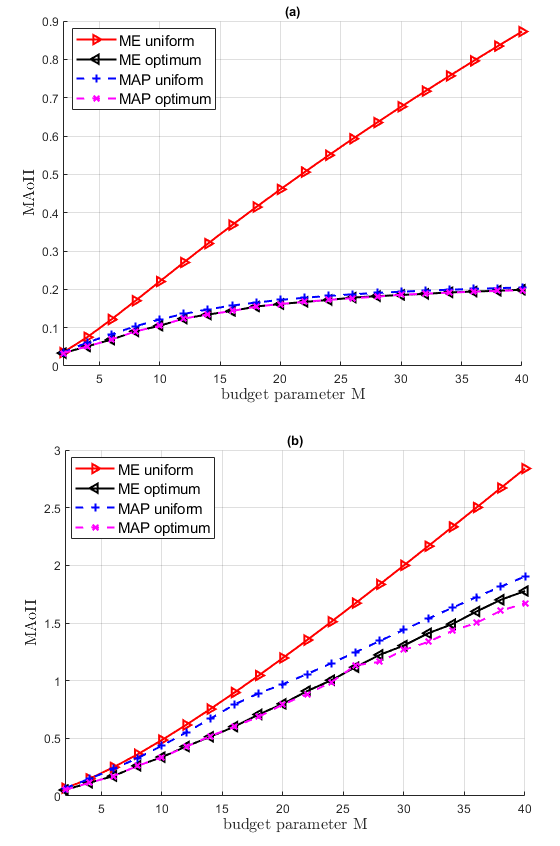}
	\caption{MAoII depicted as a function of the budget parameter $M$  employing the four pull-based policies ME uniform, ME optimum, MAP uniform and MAP optimum (a) source $A$, (b) source $B$.}
	\label{fig:Example1}
\end{figure}

In the first example, we consider the ZD channel model and two DTMC sources $A$ and $B$, with the corresponding probability transition matrices denoted by $P_A$ and $P_B$, respectively, which are set to
\begin{align}
   P_A & = \begin{bmatrix}
      0.99 & 0.01 \\
      0.20 & 0.80  
   \end{bmatrix}, \
   P_B = \begin{bmatrix}
     0.80 & 0.1 & 0.1 \\ 
     0.1 & 0.9  & 0 \\
     0.005 & 0.005 & 0.99 
   \end{bmatrix}. 
\end{align}
For both sources, the pull budget $\alpha$ is fixed to the reciprocal of an integer budget parameter $M$, i.e., $\alpha=1/M$, for which the fixed choice $g(j)=M$ ($j$ being the elapsed time since the last pull request) satisfies the sampling budget requirement with equality. Subsequently, we study the following four policies:
\begin{itemize}
    \item The {\em ME uniform} and {\em ME optimum} policies both employ the martingale estimator (ME) at the monitor. For the former policy, inter-sampling times are fixed to $M$, whereas for the latter policy, pull requests take place based on the observations optimally using the SMDP framework.
    \item {\em MAP uniform} and {\em MAP optimum} are the counterparts of ME uniform and ME optimum, respectively, where the MAP estimator replaces the ME estimator.
\end{itemize}
The MAoII is depicted as a function of the budget parameter $M$ in Fig.~\ref{fig:Example1} for the above-mentioned four policies for sources $A$ and $B$. We observe that MAP uniform substantially outperforms ME uniform in terms of MAoII for both sources, whereas MAP uniform slightly outperforms ME optimum for source $B$, whereas there does not appear to be a performance gap between the two for source $A$. On the other hand, for both sources, employing observation-dependent inter-sampling times making use of the SMDP approach is beneficial.

\begin{table}
    \centering
    \caption{Parameters used in the simulation experiments.}
    \begin{tabular}[t]{|c|c|c|}
    \hline
    $\eta$& $10^{-3}$\\ 
    \hline
    $\tau$ & $10^{-2}$\\
    \hline
    $\nu$& 0.9\\ \hline
    $\gamma$& $0.95$\\ \hline
    $\delta$  & $0.25$ \\  \hline
    numbers of hidden layers& 2\\ \hline
    nodes at hidden layers& $60$\\ \hline
    $\Delta$& 15\\ \hline
    $T_{\max}$ & $10^6$ \\ \hline
        $B$ & $10^4$ \\ \hline
    $\lambda_{\max}$ & $8$ \\ \hline
    $\eta_{\lambda}$ & $1$ \\ \hline
    \end{tabular}
    \label{tab:sim_tab}
\end{table}

\subsection{UD Channel Examples}
For the UD channel, we compare the following five policies while deploying only the MAP estimator due to its superiority over ME, which was shown for the ZD channel in the previous example.

\paragraph{Belief-CDRL} This policy refers to the CDRL algorithm described in detail in Section~\ref{sec:CDRL} for the belief-MDP formulation. The parameters used for CDRL are summarized in Table~\ref{tab:sim_tab}.

\paragraph{ZD Optimum} With this policy, we obtain the observation-dependent pull policy by assuming that the channel delay is zero. Since the policy is obtained to minimize the ZD channel, it is a sub-optimal policy for the UD channel. The ZD optimal policy obtains the best deterministic policy under the sampling rate constraint. From the nature of the discrete action space, there might be a gap between the sampling rate of the optimal deterministic policy and the actual budget. To this end, we propose the \emph{steering algorithm} for further improving ZD optimal policies, which is a non-randomized past-dependent policy. The steering algorithm mixes two deterministic policies, one with a sampling rate below the constraint and the other above it, to satisfy the constraint almost surely \cite{ross1989randomized}.  For this purpose, we denote $\lambda^i$ as the Lagrangian coefficient to obtain the policy $\pi(\lambda^i)$, and the sampling rate of this policy by $R(\lambda^i)$. Additionally, we define $R_m=\frac{1}{m}\sum_{t=1}^m a_{t}$, that stands for the current sampling rate until time $m$. Initially, we obtain two Lagrangian coefficients $\lambda^-$ and $\lambda^+$ by applying bisection search to satisfy
\begin{align}
    \lambda^-=\sup\{\lambda|R(\lambda)\leq\alpha\}, \quad \lambda^+=\inf\{\lambda|R(\lambda)\geq\alpha\}. 
\end{align}
With the steering algorithm, first the policy with the higher sampling rate, $\pi(\lambda^{+})$, is applied until $R_t$ exceeds the budget. When the budget is reached, the policy is switched to $\pi(\lambda^{-})$ until $R_t$ moves below the budget. This procedure is repeated in a loop. The steering algorithm is summarized in Algorithm~\ref{alg:steer}.

\begin{algorithm}[tbh]
    \caption{Steering algorithm}
    \begin{algorithmic}
    \renewcommand{\algorithmicrequire}{\textbf{Input:}}
    \renewcommand{\algorithmicensure}{\textbf{Initialize:}}
        \REQUIRE $\pi(\lambda^{-})$ and $\pi(\lambda^{+})$ such that $R(\lambda^{-})\leq\alpha\leq R(\lambda^{+})$
        \ENSURE $R_0=0$;
        \FOR{$t \gets 1$ to $T_{\max}$}
            \IF{$R_{t-1} < \alpha$}
            \STATE Apply the policy $\pi(\lambda^{+})$
            \ELSE
            \STATE Apply the policy $\pi(\lambda^{-})$
            \ENDIF
            \STATE $R_m \gets \frac{1}{m}\sum_{t=1}^m a_t$
        \ENDFOR
    \end{algorithmic}
    \label{alg:steer}
\end{algorithm}

\begin{figure}[thb]
    \centering
    \includegraphics[width=0.9\linewidth]{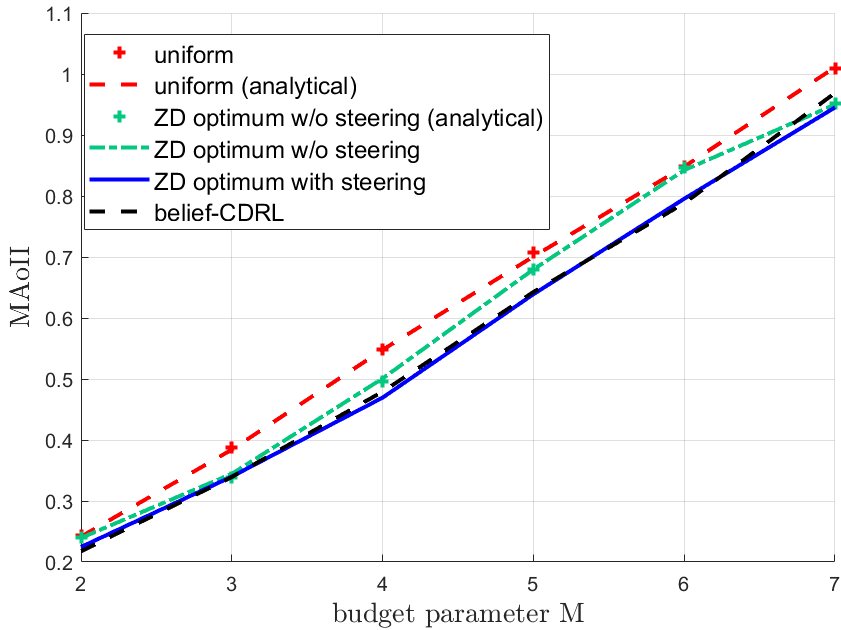}
	\caption{MAoII with respect to the budget parameter $M$ for the source given in \eqref{eq:P_C} employing uniform, ZD optimum (with and w/o steering) and belief-CDRL policies. The curves represent the simulation results, and available analytical results are marked with `$+$'.}
    \label{fig:N4}
\end{figure}

\begin{figure}[thb]
    \centering
    \includegraphics[width=0.9\linewidth]{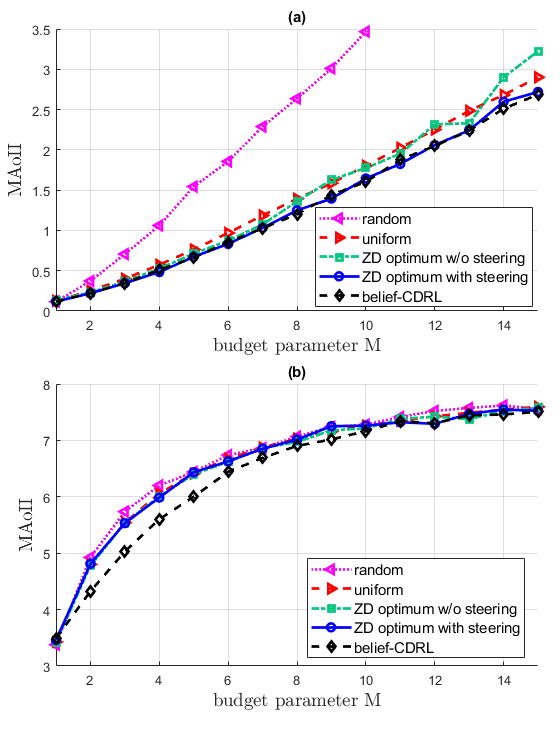} 
	\caption{MAoII with respect to the budget parameter $M$ when uniform, ZD optimum (with and w/o steering), belief-CDRL and random policies are employed a) source D in \eqref{eq:P_D} b) source E in \eqref{eq:P_E}}
    \label{fig:N5}
\end{figure}

\paragraph{Uniform} This policy is the same as the MAP uniform policy described in the previous subsection, which meets the sampling rate constraint with equality for integer $M$. The analytical results related to this policy in Fig. \ref{fig:N4} and \ref{fig:N5} are obtained using the analytical method described in Section~\ref{sec:ODpull} for this case. 
     
\begin{figure}[t]
	\centering
  \includegraphics[width=\columnwidth]{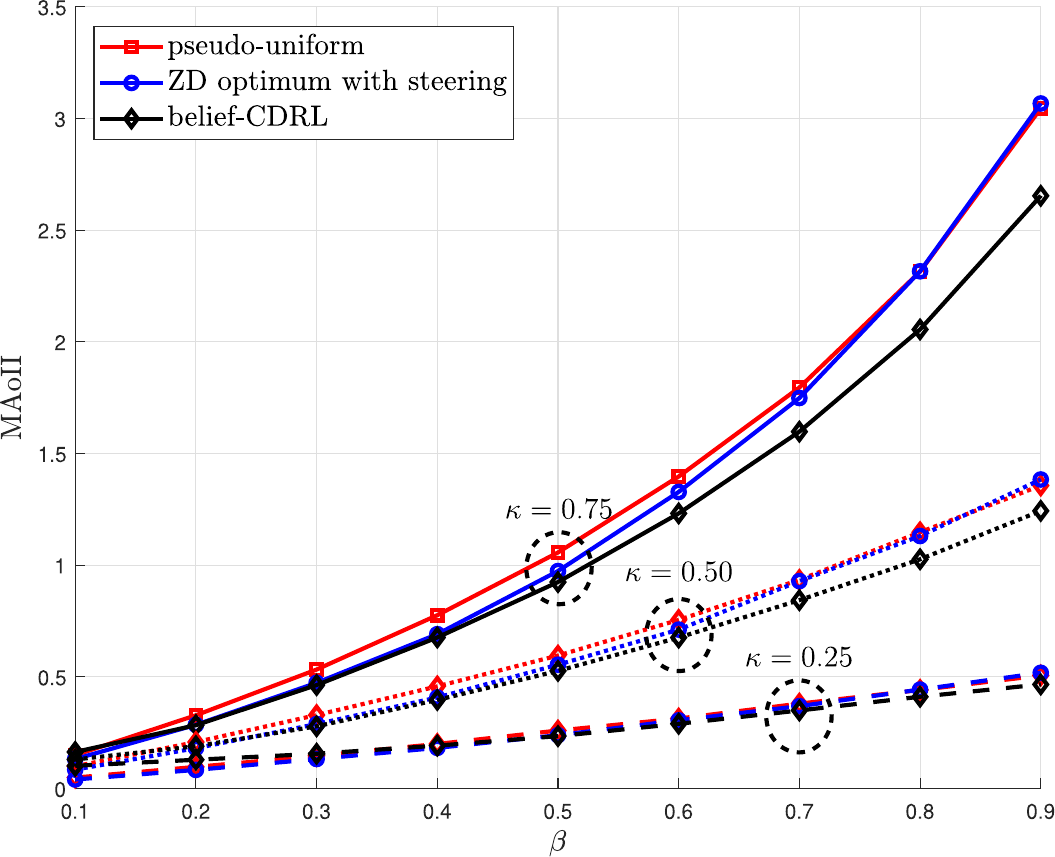}
	\caption{MAoII with respect to the parameter $\kappa$, related to the state transition frequency, for two values of the budget $\alpha$ for source G when ZD optimum with steering and belief-CDRL are employed.}
	\label{fig:fast}
\end{figure}

\paragraph{Pseudo-Uniform} For the budget values that cannot be expressed as $\alpha=\frac{1}{M}$ with an integer value $M$, the pseudo-uniform policy spreads the transmit actions as far as possible. To achieve that, the $k$th pull request is sent at the $round(k/b)$ time instance. Since this policy can no longer be expressed as an observation-dependent policy, we provide only simulation results for the pseudo-uniform policy in Fig.~\ref{fig:fast}.
     
\paragraph{Random} As another benchmark policy, we consider a policy in which, in each time slot, the monitor sends a pull request with probability $\frac{1}{M}$ when the budget parameter is $M$. Notice that this policy meets the sampling rate constraint with equality. 

Fig.~\ref{fig:N4} depicts the MAoII of the ZD optimum policy with and without steering, the belief-CDRL policy, and the uniform policy, as a function of the integer budget parameter $M$  for the DTMC source C with probability transition matrix,
\begin{align}
        P_C=\begin{bmatrix}
        0.90 & 0.05 & 0.02 & 0.03 \\
        0.10 & 0.80 & 0.05 & 0.05 \\
        0.02 & 0.01 & 0.95 & 0.02 \\
        0.10 & 0.05 & 0.10 & 0.75
        \end{bmatrix}. \label{eq:P_C}
\end{align}

For all policies, we obtain results from simulations with runtime $T_{\max}=10^6$ time slots, illustrated with solid curves. Moreover, we obtain analytical results for ZD optimum without steering and uniform policies, marked with the symbol `+'. The match between the analytical and simulation results verifies our analysis in Section~\ref{sec:ODpull}. In addition, we observe that the belief-CDRL method and the ZD optimum policy with the steering algorithm attain similar performances and both outperform the other policies for all values of $M$ we studied.

In Fig.~\ref{fig:N5}, we compare all policies for two DTMC sources, namely $D$ and $E$, with probability transition matrices,  
\begin{align}
       P_D&= \begin{bmatrix}
        0.88 & 0.05 & 0.02 & 0.03 & 0.02 \\ 
        0.05 & 0.80 & 0.05 & 0.05 & 0.05 \\
        0.01 & 0.01 & 0.95 & 0.01 & 0.02 \\
        0.01 & 0.05 & 0.01 & 0.95 & 0.02 \\
        0.05 & 0.05 & 0.05 & 0.05 & 0.80
        \end{bmatrix}, \label{eq:P_D}\\
        P_E&= \begin{bsmallmatrix}
        .240 & .075 & .094 & .090 & .108 & .042 & .107 & .097 & .091 & .057 \\
        .128 & .193 & .105 & .079 & .143 & .111 & .065 & .057 & .044 & .075 \\
        .114 & .103 & .189 & .111 & .093 & .037 & .042 & .100 & .105 & .106 \\
        .131 & .057 & .108 & .270 & .060 & .068 & .041 & .067 & .095 & .102 \\
        .097 & .086 & .041 & .096 & .227 & .117 & .132 & .115 & .047 & .042 \\
        .075 & .046 & .119 & .099 & .104 & .255 & .040 & .123 & .067 & .072 \\
        .051 & .118 & .134 & .098 & .096 & .057 & .218 & .116 & .051 & .061 \\
        .038 & .104 & .053 & .146 & .050 & .039 & .109 & .247 & .139 & .075 \\
        .065 & .089 & .090 & .126 & .066 & .102 & .083 & .074 & .186 & .118 \\
        .110 & .114 & .084 & .093 & .069 & .044 & .050 & .112 & .096 & .228
        \end{bsmallmatrix},\label{eq:P_E}
\end{align}
respectively. Notice that the source $D$ includes fewer states with state changes occurring relatively slowly compared to the process $E$. We observe that the ZD optimum with steering performs better for source $D$ than $E$, suggesting the policy obtained from the ZD model is more suitable for the UD  model when the probability of a state change in one slot is relatively lower. For both processes, the belief-CDRL policy consistently attains the lowest MAoII thanks to its capability to utilize the sufficient statistics, i.e., belief. Notice that for source $E$, state changes can occur with a larger probability than for source $D$. All policies except for the belief-CDRL policy perform similarly for this scenario. In addition, we observe that in some cases, e.g., $M=14$ for source $D$, the ZD optimum without steering performs even worse than the uniform policy since the budget constraint is not satisfied with equality when the steering algorithm is not used.

\begin{figure}[thb]
	\centering
  \includegraphics[width=\columnwidth]{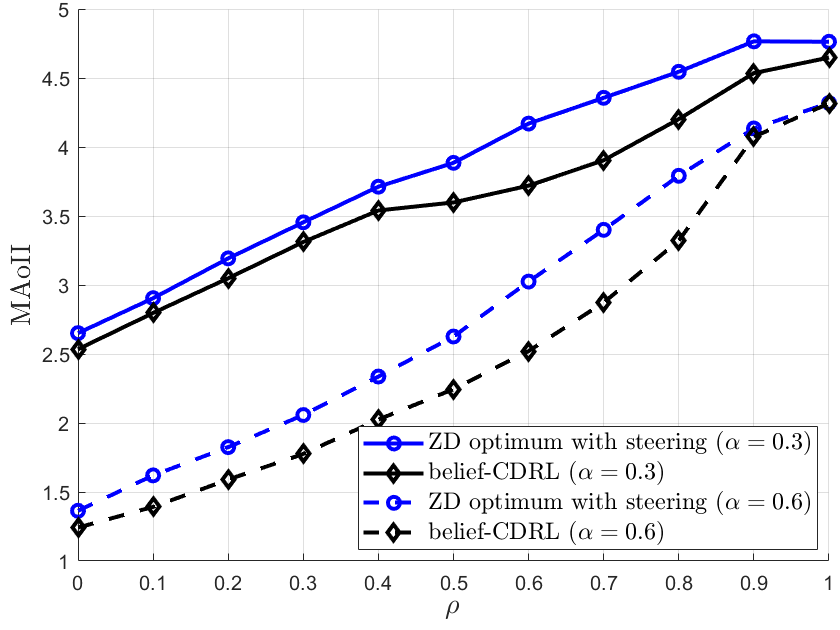}
	\caption{MAoII with respect to the heterogeneity parameter $\rho$ for two values of the budget $\alpha$ for source F when ZD optimum with steering and belief-CDRL are employed.}
	\label{fig:randomness}
\end{figure}

In the next example, we consider a DTMC source $F$ whose probability transition matrix is a linear combination of two transition matrices, namely an $8\times8$ homogeneous source with symmetric per-state behavior with transition matrix $H$,
\begin{align}
    H_{ij}=\begin{cases}
        0.55, & i=j,\\
        \frac{0.45}{7}, & i\neq j. 
    \end{cases},
\end{align}
and a heterogeneous source with highly asymmetrical states with transition matrix
\begin{align}
        R=\begin{bsmallmatrix}.121 & .090 & .167 & .131 & .156 & .151 & .015 & .169 \\ 
        .173 & .222 & .061 & .081 & .210 & .029 & .097 & .128 \\ .054 & .138 & .211 & .025 & .146 & .145 & .136 & .145 \\ 
        .115 & .072 & .102 & .236 & .144 & .153 & .096 & .082 \\ .183 & .043 & .118 & .115 & .120 & .160 & .137 & .124 \\
        .126 & .064 & .122 & .179 & .129 & .202 & .119 & .058 \\ .159 & .153 & .114 & .050 & .217 & .082 & .173 & .052 \\ 
        .092 & .144 & .115 & .030 & .093 & .125 & .161 & .240\end{bsmallmatrix},
\end{align}
with a weighting coefficient $0\leq\rho\leq1$. In particular, the transition matrix of the source is given by $P_F(\rho)=(1-\rho)H+\rho R$ with the level of heterogeneity controlled by the parameter $\rho$. In Fig.~\ref{fig:randomness}, we comparatively evaluate the policies ZD optimum with steering and belief-CDRL as a function of the heterogeneity parameter $\rho$ for two different values of the budget $\alpha=0.3, 0.6$. For both cases, we observe that when the process is close to being homogeneous (lower values of $\rho$), both policies perform similarly. This is expected since in this regime an observation-dependent policy would not be useful due to a completely symmetric source. On the other hand, as $\rho$ increases, the source becomes more heterogeneous with larger frequencies of state transitions resulting in higher MAoII for both policies. State transitions occur very rapidly in this regime, which renders both policies ineffective in this regime. However, belief-MDP exhibits higher robustness in comparison to the ZD optimum with steering.

In our final example, we construct an $8\times 8$ transition matrix $P_G(\kappa,\beta,w)$, parameterized by the three parameters $\kappa, \ \beta, \ w$, using the Markov chain construction method detailed in the Appendix. The steady-state distribution of the process is solely controlled by the parameter $0<\beta\leq1$, and is spread linearly from $\frac{\beta}{N}$ to $\frac{\beta}{N}+\frac{2(1-\beta)}{N}$. Thus, $\beta=1$ generates a fully symmetric process with $\pi_n=\frac{1}{8}, \ n=\{1,\dots,N\}$. On the other hand, maximum separation among the stationary probabilities is achieved as $\beta \rightarrow 0$. In addition, the speed of the Markov chain, denoted by $s$, is defined as the average frequency of state transitions per unit time and is given as $s=\beta\kappa$, so that as $\kappa\rightarrow 1$, the Markov chain changes its states more frequently. In this example, we fix the parameters $w=\frac{1}{49}\begin{bmatrix}
    4 & 5 & 6 & 7 & 8 & 9 &10
\end{bmatrix}$, $\alpha=0.6$ and illustrate the performance of the pseudo-uniform, ZD optimum with steering and belief-CDRL policies for $\kappa=\{0.25,0.5,0.75\}$ and varying $\beta$ values in Fig.~\ref{fig:fast}. First, we observe that MAoII values of policies are monotonically increasing with $\kappa$ and $\beta$. This verifies the negative correlation between the speed of the process increases, the ability of remote estimating it. We additionally observe that as the steady-state distribution of the process becomes more symmetric, ZD optimum with the steering policy converges to the pseudo-uniform policy regardless of the $\kappa$ values. On the other hand, the policy belief-CDRL has a consistent performance gain compared to pseudo-uniform, which is around $\%10$ in this example. 

\section{Conclusions} \label{sec:conclusions}
We considered the minimization of the mean age of incorrect information (MAoII) for a pull-based remote estimation system under a limited sampling budget when the source is a DTMC process whose characterization is known in advance.  For the estimation rule, we use martingale estimators (ME) along with maximum a posteriori (MAP) estimators. When the channel from the source towards the monitor is a perfect channel, i.e., zero-delay (ZD) channel, we formulate the unconstrained version of the problem as an SMDP and use policy iterations to obtain optimum deterministic policies. Subsequently, we use bisection search to obtain the solution to the constrained problem using the Lagrange method. However, deterministic policies may not meet the budget constraints with equality. For this purpose, we deviate from deterministic policies by proposing a steering algorithm which mixes two deterministic policies to further improve the MAoII performance while satisfying the budget constraints almost equally. For the unit-delay (UD) channel, the monitor does not have perfect information on the source's state or the AoII process at any time. Therefore, the SMDP method used for the ZD channel cannot be used for this case. We propose a constrained reinforcement learning (CRL) called belief-CRL by which the monitors send belief-dependent pull requests for unit delay channels. The proposed belief-CRL method is shown to outperform various benchmark policies in all the examples we studied, and the gain attainable by belief-CRL is more apparent for faster or heterogeneous DTMC source processes. Extension to more general channels is left for future work. We additionally suggest that the proposed belief-dependent control approach has the potential to be used in other related information freshness problems.

\appendix[Construction of Source DTMCs with Controlled Asymmetry and Speed]
To assess the performance of the proposed policies over a broad range of source dynamics, we construct a parametric family of $N$-state discrete-time Markov chains (DTMCs), $X_t$, $t\in\{1,2,\dots\}$, with probability transition matrix $P$ in which two structural properties of the information source are controlled independently and explicitly: (i)~the \emph{asymmetry} of the stationary distribution across states, controlled by parameter $\beta$, and (ii)~the \emph{speed} of the chain, defined as the average frequency of actual state changes per time slot, controlled by parameter $\kappa$. In addition, we consider a weight vector $w=(w_1,\ldots,w_{N-1})$  that describes the state transition probability, which is fixed in a circular way. In other words
\begin{align}
    \mathbb{P}(X_{t+1}=j|X_t=i,i\neq j)=w_{j\ominus i}, \label{eq:omega}
\end{align}
where $j\oplus i \triangleq \big((j+i-1) \bmod N\big)+1$ and $j\ominus i \triangleq \big((j-i-1) \bmod N\big)+1$ denote cyclic addition and subtraction over $N$, respectively.

\subsection{Target Stationary Distribution}
Let the state space be $\mathcal{S}=\{1,\ldots,N\}$, $N\ge 2$. We impose a \emph{linearly spaced} stationary distribution
$\pi=(\pi_1,\ldots,\pi_N)$,
\begin{align}
  \pi_j = \pi_1 + (j-1)\Delta, \quad j=1,\ldots,N, \label{eq:linear-pi}
\end{align}
parameterized by a single asymmetry parameter $\beta\in(0,1]$ via
\begin{align}
  \pi_1 = \frac{\beta}{N}, \quad \Delta = \frac{2(1-\beta)}{N(N-1)}, \label{eq:pi1-delta}
\end{align}
where $\Delta$ follows from the normalization $\sum_{j=1}^{N}\pi_j = 1$. The parameter $\beta$ interpolates between the two extremes: $\beta=1$ gives rise to the uniform (fully symmetric) distribution $\pi_j=1/N$, whereas $\beta\to 0$ yields maximum asymmetry with $\pi_1\to 0$ and $\pi_N\to 2/N$. We also define $\pi_{\min}=\pi_1$.

\subsection{Cycle-Based Construction of the Transition Matrix}
Rather than solving an inverse eigenvector problem for a transition matrix admitting $\pi$ in \eqref{eq:linear-pi} as its stationary vector, we build the chain constructively as a convex combination of
\emph{directed-cycle} chains, in the spirit of the cycle representation of Markov processes \cite{kalpazidou2006}. For each step size $i\in\{1,\ldots,N-1\}$, we define the slice $Q^{(i)}$ in which we have a forward jump from state $j$ by $i$ positions (cyclically) or self-loops:
\begin{align}
  Q^{(i)}_{j,j\oplus i} = q_j,
  \quad
  Q^{(i)}_{j,j} = 1-q_j,
  \quad
  q_j \triangleq \kappa\frac{\pi_{\min}}{\pi_j},
  \label{eq:slice}
\end{align}
where $\kappa\in(0,1]$ is the speed parameter to be described below. Since $\pi_j\ge\pi_{\min}$ for all $j$, we have $q_j\le\kappa\le 1$, thus each $Q^{(i)}$ is a valid stochastic matrix.

The key property of \eqref{eq:slice} is that the probability flow out of every state is constant along each directed cycle:
\begin{align}
  \pi_j q_j = \kappa\pi_{\min}, \quad \forall j\in\mathcal{S}.
  \label{eq:constant-flow}
\end{align}
Consequently, for every $j$, the stationary flow entering $j$ (from state $j\ominus i$) equals the stationary flow leaving $j$, and the global balance equations \cite[Ch.~1]{norris1997}
\begin{align}
  \sum_{k\in\mathcal{S}} \pi_k Q^{(i)}_{kj}
  & = \pi_j(1-q_j) + \pi_{j\ominus i} q_{j\ominus i} \\
  & = \pi_j - \kappa\pi_{\min} + \kappa\pi_{\min} 
  = \pi_j
\end{align}
hold for every slice, i.e., $\pi Q^{(i)} = \pi$ for all $i$. 

The final transition matrix is the mixture
\begin{align}
  P = \sum_{i=1}^{N-1} w_i Q^{(i)},
  \quad
  w_i \ge 0,\quad \sum_{i=1}^{N-1} w_i = 1,
  \label{eq:mixture}
\end{align}
where the weight vector $w=(w_1,\ldots,w_{N-1})$ shapes the jump size distribution without affecting the stationary vector. This can be seen from the linearity of the balance equations $\pi P=\pi$ for any choice of $w$ in \eqref{eq:mixture}. The chain $P$ is irreducible if and only if $\gcd\!\big(N,\{i: w_i>0\}\big)=1$ (in particular, whenever $w_1>0$), and it is aperiodic whenever at least one self-loop probability in
\eqref{eq:slice} is positive, i.e., unless $\kappa=\beta=1$ and a single step size is used \cite[Ch.~1]{levin2017}. 

\subsection{Speed of the Chain}
We define the \emph{speed} $s$ of a DTMC as the long-run expected fraction of time slots in which an actual state change occurs, i.e., the average frequency of state transitions per unit time,
\begin{align}
  s \triangleq \sum_{j=1}^{N} \pi_j\big(1 - P_{jj}\big).
  \label{eq:speed-def}
\end{align}
Since every off-diagonal move in \eqref{eq:slice} yields a transition from state $j$ to state $j\oplus i$ for $1\le i\le N-1$, and since the jump probability out of state $j$ under the mixture \eqref{eq:mixture} equals
$\sum_i w_i q_j = q_j$ irrespective of $w$, the speed has the closed-form
\begin{align}
  s = \sum_{j=1}^{N} \pi_j\frac{\kappa \pi_{\min}}{\pi_j}
    = N \kappa \pi_{\min}
    = \kappa \beta.
  \label{eq:speed}
\end{align}
Thus, the parameter $\kappa$ acts as a uniform speed multiplier, i.e., for a fixed asymmetry level $\beta$, the speed scales linearly in $\kappa$, independently of $N$ and of the mixture weights $w$. 

\bibliographystyle{ieeetr}
\bibliography{bibl}

\end{document}